\documentclass[prb,superscriptaddress,twocolumn,showpacs,showkeys,amsmath,
amssymb]{revtex4}

\usepackage{graphicx}
\usepackage{dcolumn}
\usepackage{bm}

\usepackage{pstricks}
\usepackage{pst-all}
\usepackage{psfrag}

\usepackage{array}

\usepackage{epsfig}

\newcommand{\f}[2]{{\ensuremath{\mathchoice%
	{\dfrac{#1}{#2}}
        {\dfrac{#1}{#2}}
	{\frac{#1}{#2}}
	{\frac{#1}{#2}}
	}}}

\newcommand{\U}[1]{\ensuremath{\mathrm{~#1}}}

\usepackage{color}

\newcommand{\ket}[1]{\ensuremath{| #1 \rangle}}

\newcommand{\kb}[2]{\ensuremath{|#1\rangle \langle #2|}}
\newcommand{\lr}[1]{\ensuremath{\langle #1 \rangle}}

\newcommand{\op}[1]{\ensuremath{\mathbf{\hat{#1}}}}

\begin{document}

\preprint{APS/123-QED}

\title{Quantum coherence and carriers mobility in organic semiconductors}

\author{J.-D. Picon}
\affiliation{%
Laboratory of Optoelectronics of Molecular Materials\\
\'Ecole polytechnique f\'ed\'erale de Lausanne\\
LOMM / IMX / STI, Station 3\\
CH-1015 Lausanne} 

\author{M. N. Bussac}
\affiliation{%
Center for Theoretical Physics, CNRS\\
\'Ecole polytechnique,\\
F-91128 Palaiseau Cedex
} 

\author{L. Zuppiroli}
\email[Author to whom the correspondance should be
adressed, e-mail :]{libero.zuppiroli@epfl.ch}
\affiliation{%
Laboratory of Optoelectronics of Molecular Materials\\
Swiss Federal Institute of Technology -- Lausanne\\
LOMM / IMX / STI, Station 3\\
CH-1015 Lausanne} 



\date{\today}

\begin{abstract}

We present a model of charge transport in organic molecular semiconductors based on the effects of lattice fluctuations on the quantum coherence of the electronic state of the charge carrier. Thermal intermolecular phonons and librations tend to localize pure coherent states and to assist the motion of less coherent ones. Decoherence is thus the primary mechanism by which conduction occurs. It is driven by the coupling of the carrier to the molecular lattice through polarization and transfer integral fluctuations as described by the hamiltonian of Gosar and Choi. \cite{gosar} Localization effects in the quantum coherent regime are modeled via the Anderson hamiltonian with correlated diagonal and non-diagonal disorder \cite{mnb} leading to the determination of the carrier localization length. This length defines the 
coherent extension of the ground state and determines, in turn, the diffusion range in the incoherent regime and thus the mobility. The transfer integral disorder of Troisi and Orlandi \cite{troisi} can also be incorporated. This model, based on the idea of decoherence, allowed us to predict the value and temperature dependence of the carrier mobility in prototypical organic semiconductors that are in qualitative accord with experiments.

\end{abstract}

\pacs{ 72.80.Le, 71.23.An, 33.15.Kr
}
\keywords{ organic semiconductors, localized states, polarization,  electronic
polaron}

\maketitle

\section{Introduction}

Molecular electronics is a field that is rapidly gaining importance because of its potential in producing a new breed of plastic organic devices.  A physical quantity that is critical to their operation is the charge carrier mobility in the organic molecular semiconductor that forms the active layer in these devices. It is therefore not surprising that along with this resurgence of interest in molecular electronics in recent years, charge transport mechanisms in organic semiconductors are once again at the forefront condensed-matter physics research. Consequently, important debates in this field have been rekindled recently to which we would like to contribute by the present work.


Experimental observations of "band-like" charge transport in single crystalline organic molecular semiconductor have previously been observed in stilbene \cite{kostin}, perylene \cite{karl}, acenes \cite{warta,karl2} or rubrene. \cite{podzorov,stassen} Hole mobilities of 1 to 10 cm$^2$/V.s at room temperature were reported. In many cases, these mobilities were shown to decrease with increasing temperatures. At low temperatures, values of a few hundred cm$^2$/V.s were even seen in a "time-of-flight" measurement. \cite{karl} Earlier theoretical explanations of these results invoked rigid band models to account for the power-law temperature dependence of the mobility. \cite{warta} Later on, the electron-phonon interaction became widely recognized as the key factor that determines the mobility, leading to the development of several polaronic models of charge transport in these materials.  \cite{gruhn,coropceanu,silvafilho,mastorrent,kenkre,hannewald, hannewald2}


The first proposed polaronic models have emphasized the influence of the molecular character of organic molecular crystals on charge transport. One approach calculated the reorganization energies on a single molecule and used the Marcus theory of charge transfer \cite{gruhn,coropceanu,silvafilho,mastorrent} to determine the mobility. Another \cite{kenkre, hannewald, hannewald2} stressed the importance of low energy phonons and librations and derived an average mobility from the Kubo formula.  In their recent works, Troisi and Orlandi criticized both these approaches based on polaron theories. They have suggested that the formation of a small polaron with the charge localized on a single molecule is unlikely \cite{troisi2} because the nuclear reorganization energy in these crystals is comparable to the average intermolecular charge transfer while the more general approach based on intermolecular vibrations suffers some problems of averaging.\cite{troisi} Subsequently, these models were extended to describe the effects of thermal disorder and thermal fluctuations on charge transport. In this case, low energy phonons and librations are able both to localize the charge and to drive its diffusion in the lattice. In a previous work, \cite{mnb} we implemented the idea of phonons acting as a source of disorder and localizing the charge using an Anderson hamiltonian with the carrier coupled to low energy intermolecular phonons and librations through the fluctuations of the polarization energy. On the other hand, the model of Troisi and Orlandi focused on the effects of fluctuations of the transfer integral which are considered to be large enough to localize and then move the carrier in the lattice. \cite{troisi} We note that these theoretical considerations \cite{mnb,troisi} have been introduced in the pioneering work of Gosar and Choi \cite{gosar} on the determination of the mobility of an excess charge in a molecular crystal.

%


The present work focuses on quantum coherence effects on charge transport in organic semiconductors that has been ignored so far because of the technical difficulty of treating decoherence in a system with strong electron-phonon interactions. We calculate the carrier localization length resulting from the fluctuations of the electronic polarization energy. The fluctuations of the transfer integrals treated in Ref. \onlinecite{troisi} can also be included without problem. This localization length allowed us to determine the carrier diffusion coefficient leading to the prediction of the mobility of charge carrier in organic molecular crystals. We find that these results can explain the main trends of time of flight mobility measurements performed in pure organic semiconductor single crystals.

\section{The general framework of the conduction model}

Low energy phonons and librations play an ambivalent role in limiting charge
carrier transport in molecular semiconductors. Depending on the quantum
coherence of the electronic state of the carrier, thermal phonons will tend to
localize pure coherent states or assist the motion of less coherent ones. As
long as the carrier electronic states keep their quantum coherence, they are
essentially localized by thermal disorder within a localization length $L$, which
defines the coherent extension of the ground state. At longer time scales, thermal
fluctuations cause decoherence and assist the diffusive motion of the carrier.
This process is driven by the coupling of the carrier to the molecular lattice,
through polarization fluctuations and transfer integral fluctuations as
described by Gosar and Choi.\cite{gosar}
The localization length $L$ determines the diffusion range and thus the
mobility. To determine this quantity in the quantum coherent regime, we first
introduce a transfer matrix formalism in two dimensions depicting the quantum
interference processes in the plane of high conduction of these organic
semiconductors. The hamiltonian for this quantum model is the Anderson
hamiltonian \cite{johansson} with correlated diagonal and non-diagonal disorder
established in Ref. \onlinecite{mnb}, and which is extended in the present work
by adding the purely diagonal transfer integral disorder computed in Ref.
\onlinecite{troisi}. The role of spatial and energetic correlations will be
studied with particular emphasis for the case of polarization fluctuations. 
For the sake of concreteness, we shall apply the results derived here to widely
studied molecular single crystals such as the acenes or rubrene. 

The lattice
dynamics in these materials have been investigated extensively. Inelastic
neutron scattering in naphtalene by Natkaniec et al. \cite{natkaniec} revealed
twelve intermolecular phonon branches in crystalline acenes. Structural
refinements using a molecular dynamics model yielded a satisfactory fit to the
measured phonon dispersion curves. More recently, some of these modes along with
their relative electron-phonon coupling constants were obtained by a density
functional theory scheme (DFT/LDA) applied to the acene series from naphtalene
to tetracene. \cite{hannewald2} Both these calculations converge on the
existence
of a vibrational band centered near 50 cm$^{-1}$ with 3 acoustic modes, 3
optical modes, and 6 librations. The directional average of the root-mean-square
amplitudes of the translational vibrations of the molecules is $0.17\text{\AA}$ at room
temperature in anthracene as obtained from the X rays measurements of
Cruickshank. \cite{cruickshank} This work also showed a typical librational
amplitude of 3 or 4 degrees at room temperature.
The largest hole transfer integral between adjacent sites in pentacene, the
prototype acene compound, is about 1000 cm$^{-1}$  (0.12 eV) as determined
independently in Refs. \onlinecite{troisi2} and \onlinecite{cheng}. Thus, the
thermal motion can essentially
be considered as static within the hole residence time of the order of 0.1 ps. More details on time scales are given in Appendix \ref{time}.
This is enough time to build a large quantum coherence into the system based on
quantum interference in the Anderson weak localization regime.
For times longer than 1 ps, the coupling of the extra charge with the
intermolecular phonons reservoir now becomes the source for decoherence.
\cite{pastawski} In
this case, we treat the charge motion within an adiabatic classical diffusion
approximation, as the charge follows the motion of the low energy phonon
wavepacket. The characteristic length of this diffusion is just the localization
length determined in the short time scale coherent regime. 
Intramolecular vibrations play a negligible role in the decoherence. In fact,
according to Ref. \onlinecite{coropceanu}, there is essentially one mode that
contribute to the
reorganization energy of the positively ionized molecule. In pentacene, this
corresponds to vibrational modes at 1340 cm$^{-1}$. This
mode is
too fast to localize the charge. Its main effect is to renormalize the
transfer integral as outlined in our calculation presented previously in
Appendix C of Ref. \onlinecite{hocine}. That calculation yielded a further
reduction of the
bare transfer integral by a factor of 0.75. 

\section{The calculations of the localization lengths}

\subsection{The hamiltonian}

On each site of a real molecular crystal at finite temperature, the molecular positions and angles fluctuate with respect to their values in the perfect crystal. The carrier motion in the lattice is determined by its interaction with these fluctuations. 
In order to construct the hamiltonian that properly accounts for these interactions, a clear distinction must be made between fast and slow interactions with respect to the transfer time $h/J^0$, where $J^0$ is the larger transfer integral in the conducting plane of the molecular semiconductor. This classification of the interactions according to their time scale is described in Appendix \ref{time}: we show that the interaction with the electronic polarization and with intramolecular vibrations are fast, while the interaction with intermolecular phonons and librations is slow. Fast interactions can be averaged; they just renormalize the parameters of the hamiltonian.\cite{hocine} On the contrary, slow or static interactions have to be included explicitly in the hamiltonian.

This is the case for intermolecular phonons, which as we mentioned earlier, form a band around $50 \text{cm}^{-1}$ in acenes. These displacement fields on each crystal site can be considered as uncorrelated because of the nearly degenerate twelve phonon modes of acenes : many low-frequency phonons can also be treated as independent random displacements $\delta r_j$ and $\delta \theta_j$ on each site $j$. These lattice fluctuations as shown below, induce, in turn, site energy fluctuations $\delta_j$ and transfer integral fluctuations $\delta'_{j, j+h}$. These random variables are characterized by mean square values $\delta$ and $\delta'$ respectively. We can now start the construction of the hamiltonian.

In a previous paper \cite{mnb}, the long range Coulomb hamiltonian describing
the Coulomb polarization induced by a charge in the molecular lattice is mapped onto a
short range tight binding hamiltonian containing the polarization energy
$E_{\text{p}}(n)$ on each molecular site $n$ and the renormalized transfer
integral $\tilde{J}_{n,n+h}$ which couples two adjacent sites $n$ and $n+h$.

\begin{equation}
\op{H}=\sum_{n} E_p(n)\kb{n}{n}-\sum_{n,n+h}\tilde{J}_{n,n+h}\kb{n}{n+h}
\label{eq-1}
\end{equation}

In the case of a perfect crystal, the polarization energy is uniform and shifts
the ground state uniformly by about 1 eV as observed experimentally in acenes.
\cite{pope, song} Consequently, the bare bandwith is significantly narrowed,
regardless of the temperature, both by the polarization cloud and the
intramolecular phonon cloud which dress the charge. The reduction factors
of the bare integral $J$ have been calculated in Ref. \onlinecite{hocine} for
pentacene to be 0.79 for polarization and 0.75 for intramolecular phonons.

The situation changes greatly when intermolecular thermal disorder enters the system.  Thermal disorder on each site is parameterized by six Gaussian random
variables attributed to each translational and librational degree of freedom of
the molecules.

We designate the position fluctuations $\delta\vec{R_j}=\vec{r_j}-\vec{r_j}^0$
and the angular fluctuations are represented by
$\delta\vec{\theta_j}=\vec{\theta_j}-\vec{\theta_j}^0$. Following
Ref. \onlinecite{mnb}, the polarization energy in the presence of thermal
disorder
can be written at each site $n$ as

\begin{equation}
E_{\text{p}}(n) = E_{\text{p}}^0(n) + \delta_n
\label{eq-3}
\end{equation}

where $\delta_n=\sum_{j\neq n} ({\partial
E_{\text{p}}(n)}/{\partial\vec{r_j}})\,(\delta \vec{r_j}-\delta \vec{r_n})+
({\partial E_{\text{p}}(n)}/{\partial\vec{\theta_j}})\,(\delta
\vec{\theta_j}-\delta
\vec{\theta_n})$. The renormalization factor of the transfer integral at each
site
due to polarization effects is given by \cite{mnb} 

\begin{equation}
\frac{\tilde{J}_{n,n+h}}{J_{n,n+h}}=\exp\frac{\delta_n+\delta_{n+h}}{\Delta}
\end{equation}

where $\Delta$, an energy of the order 1 eV, defines the importance of
renormalization. \cite{mnb} It is also possible to introduce in the hamiltonian the
fluctuations of the bare transfer integrals, $J_{n,n+h}$, following the
model of Troisi and Orlandi \cite{troisi} as
\begin{equation}
J_{n,n+h}(\vec{r}_{n+h}-\vec{r}_,\vec{\theta}_{n+h}-\vec{\theta}_n) =
J_{n,n+h}^0+\delta_{n,n+h}'
\label{eq-4}
\end{equation}

\bigskip

where $\delta_{n,n+h}'=\vec{\alpha}(\delta\vec{r}_{n+h}+\delta\vec{r}_n)+\vec{
\gamma}
(\delta\vec{\theta}_{n+h}-\delta\vec{\theta}_n)$.
Then hamiltonian \ref{eq-1} becomes an Anderson hamiltonian with correlated
diagonal and non diagonal disorder and includes both polarization fluctuations, $\delta_n$,
and transfer integral fluctuations, $\delta_{n,n+h}'$.

\begin{widetext}

\begin{equation}
\op{H}=\sum_{n} (E^0_{\text{p}}(n) +
\delta_n)\kb{n}{n}-\sum_{n,n+h}(J_{n,n+h}^0+\delta_{n,n+h}')\exp\frac{\delta_n+\
\delta_{n+h}}{\Delta }\kb{n}{n+h}
\label{eq-5}
\end{equation}

\end{widetext}

In principle, the integral energy fluctuations $\delta_{n,n+h}'$ and the
polarization fluctuations $\delta_n$ are correlated because they are derived
from the same thermal displacement field
$(\delta\vec{r}_n,\delta\vec{\theta}_n)$.
It is possible that the complexity of the displacement
fields dilutes the effects of these correlations. 
Troisi
and Orlandi have studied  the effects of $\delta_{n,n+h}'$ alone. Here we shall
focus on the polarization fluctuations $\delta_n$, which appear both in the
diagonal and the nondiagonal part of the Anderson hamiltonian.
Thus in the following  we shall set $\delta_{n,n+h}'=0$ in the hamiltonian of Eq. \ref{eq-5}. We are left with the random variable $\delta_n$ the distribution of which has a root mean square value $\delta$.

Polarization fluctuations effects are long ranged. Indeed the polarization cloud extends over many molecules as the contribution of the induced dipoles decrease like the reciprocal distance to the carrier $(1/r)$. It is thus important in our calculations to
explore the amplitude of the correlations. This has been done in Appendix \ref{correlations}.
The result
is that spatial correlations 
due to these disordered dipoles are much shorter range. Thus
 the correlations between the polarization energy of two adjacent sites can be neglected. Nevertheless, diagonal and nondiagonal disorder are correlated through the renormalization factor $\Delta$.
The disorder parameters which constitute the starting point of the transfer
matrix calculation are presented in table \ref{tab-1}. The root mean square values $\delta$ of the polarization fluctuations have been calculated as follows. For a given distribution of lattice displacements that we know from the experimental work of Cruickshank\cite{cruickshank}, the polarization energy has been calculated according to our previous papers.\cite{mnb,hocine} Then by varying the number of samples, the distribution of $\delta_n$ can be reconstructed. We have checked that it is gaussian with a root mean square energy $\delta$.

\begin{table}
\caption{\label{tab-1} Thermal energetic disorder $\delta$ resulting from the
translational, librational and both translational and librational lattice fluctuations is the root mean square value of the disorder distribution. The renormalization factor $\Delta$ independent of the type of
disorder (see Eq. \ref{eq-5}) is related essentially to the HOMO-LUMO gap\cite{mnb}}
\begin{ruledtabular}
\begin{tabular}{ccc}
Disorder &$\delta $ (meV) & $\Delta$ (eV)\\\hline
$0.1$ \AA& 42.1  & $\simeq 0.4$\\
3 degrees & 9.4 & $\simeq 0.4$\\
3 degrees + $0.1$ \AA&44.9 & $\simeq 0.4$\\
\end{tabular}
\end{ruledtabular}
\end{table}

%
%

\subsection{The localization length}

The hamiltonian of the carrier coupled to slow lattice fluctuations of Eq. \ref{eq-5} is of the Anderson type (with correlated diagonal and nondiagonal disorder). It is well known that such types of hamiltonian lead to Anderson localization: the carrier quantum state results from the interference process between the wavelets scattered at each site; this process weakens the forwards scattering in favor of the backwards one. Such a problem cannot be treated by averaging the energy distribution because the coherence of the quantum state, expressed by the phases of the wave function at each molecular site, should be preserved. The only transport model which is completely quantum is the transfer matrix formalism \cite{unge, kramer}, which determines, in amplitude and phase, the carrier transmission and reflexion coefficients in a disordered lattice. These coefficients cannot be directly averaged. Only the Lyapunov of the distribution obeys a central limit theorem and can be averaged. \cite{unge, kramer,unge2} We used the transfer matrix formalism to calculate the localization length of the carrier. In two dimensions this method can be applied to a long strip.

 \begin{figure}[!h]
\begin{center}
\epsfig{file=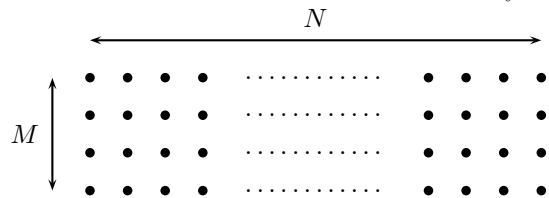}
%
%
%
%
%
%
%
%
%
%
%
%
\end{center}
\caption{Two-dimensional strip of width $M$ and length $N$.}
\label{fig-5}

\end{figure}

We consider a 2D lattice array system of size $N\times M$ sites as illustrated
in Figure \ref{fig-5}. Each site $(n,m)$ corresponds to one molecule and is 
characterized by an energy $\delta_{nm}$ from a gaussian
distribution. The ground state wavefunction $\ket{\psi}$

\begin{equation}
\ket{\psi} =\sum{}{} a_{n,m}|n,m\rangle
\end{equation}

\noindent is solution of the Schr\"odinger equation
\begin{equation}
  \op{H}\ket{\psi}=E\ket{\psi}
\label{eq-7}
\end{equation}

We impose periodic boundary conditions and only take into account
next neighbor interactions.
Then equation \ref{eq-7} can be written explicitly as,
\begin{widetext}
  \begin{eqnarray}
   & \langle n,m|(\hat{H}-E)\ket{\psi}=0 \\\nonumber
   & \Leftrightarrow (\delta_{n,m}-E) a_{n,m} + J_{n,m}^{n,m+1}
    a_{n,m+1}+J_{n,m}^{n,m-1} a_{n,m-1} +
    J_{n,m}^{n+1,m} 
    a_{n+1,m}+J_{n,m}^{n-1,m} a_{n-1,m} =0 
  \end{eqnarray}
\end{widetext}

Using $A_n=(a_{n,1},\dots,a_{n,M})$ then the former equation
becomes,
\begin{equation}
A_{n+1}=M_{n}A_{n}+\tilde{M}_{n-1}A_{n-1}
\label{eq-9}
\end{equation}

\noindent with $M_n$ and $\tilde{M}_n$ given in Appendix
\ref{matrices}. This equation can be recast into the transfer matrix formalism
as,

\begin{equation}
\left(
\begin{array}{cc}
A_{n+1}\\
A_{n}\\
\end{array}
\right)
=\left(
\begin{array}{cc}
M_{n+1}&\tilde{M}_{n}\\
\mathbb{I}&\mathbb{O}\\
\end{array}
\right)
\left(
\begin{array}{cc}
A_{n}\\
A_{n-1}\\
\end{array}
\right)
=T_n\left(
\begin{array}{cc}
A_{n}\\
A_{n-1}\\
\end{array}
\right)
\end{equation} 

We define $\tau_N=\prod_{1}^{N}T_n$ then

\begin{equation}
\left(
\begin{array}{cc}
A_{n+1}\\
A_{n}\\
\end{array}
\right)
=\tau_n \left(
\begin{array}{cc}
A_{1}\\
A_{0}\\
\end{array}
\right)
\end{equation}

\noindent For the limiting case of $n$ going to infinity, $\tau_n$ exposes the
asymtotic behaviour
of the ground state wavefunction $\ket{\psi}$. 
According to Oseledec's theorem\cite{oseledec},
the following limit yields the quantity $\gamma_i$, called $i$-th Lyapunov
exponent where $v_i$ is $i$-th eigenvalue of $\tau_n$. 

%
%
%

\begin{equation}
\lim_{n\rightarrow\infty}\frac{\ln(v_i)}{n}=-\gamma_i
\end{equation}

\noindent is well defined so that when $n$ increases , $v_i$ is simply

\begin{equation}
v_i\sim\text{e}^{-n\gamma_i}
\end{equation}

Now, $\gamma_i>0$ describes the exponential rate of
decrease  of the components of the ground state $\ket{\psi}$. Then, any
solution of the
Schr\"odinger equation will decrease faster than the minimum rate
given by 

\begin{equation}
\gamma=\min_{i}(\gamma_i)
\end{equation}

$L={1}/{\gamma}$ is called localization length and represents a measure of
the extension of the ground state wavefunction.

To calculate the localizaton length, we sorted $2N$ random vectors and  
calculated their
evolution via the transfer matrix process. Actually, if we start from a
vector $V_0$, after $n$ steps, its image $V_n$ is dominated by the
largest Lyapunov exponent. To obtain the smallest one, that corresponds to the
localization length, we used an orthogonal normalization \cite{benettin} process
applied to $2N$ random vectors in order to get rid of the contribution of the
$2N-1$ first Lyapunov exponents that screen the one we look after.

%

\begin{figure}[!h]
\begin{center}
\psfrag{w}[][]{Reduced localization length,  $L(M)/M$} 
 \includegraphics{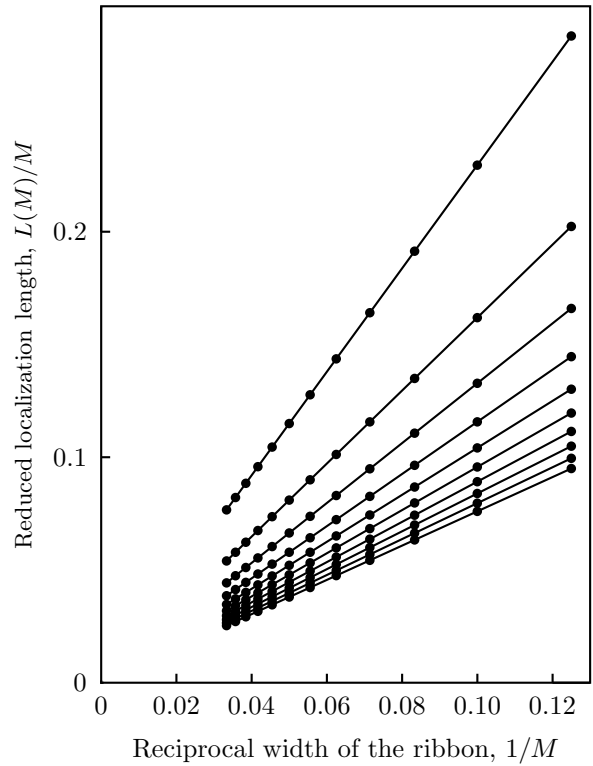}
%
%
%
%
%
%
%
%
%
%
%
%
%
%
%
%
%
\end{center}
\caption{Localization length in a strip of finite width $L(M)$. Each curve corresponds to a different energy in the band and a
disorder of $0.5J$. When $1/M$ tends to 0, the localization length for a given
energy corresponds to the asymptotic slope of the corresponding curve.}
\label{fig-4}

\end{figure}

These Lyapunov exponents were calculated for arrays of finite width $M$. We have to use a finite-size rescaling to extract the actual localization length $L$ from the localization length $L(M)$ for finite values of $M$. According to
Ref. \onlinecite{mackinnon}, as the diagonal parameter $\delta/\tilde{J}$ is
lower than 4, we identify in our 2D system the localization length $L$ of the
infinitely wide array by taking the limit

\begin{equation}
\lim_{M\rightarrow\infty}L(M)
\end{equation}

\noindent where $L(M)$ is the localization length obtained for an array of
width $M$. Figure \ref{fig-4} shows the values of $L(M)/M$ versus
$1/M$ for different energy positions in the bandwidth calculated for
$\Delta=0.4$ as established in Ref. \onlinecite{mnb} for an angular
disorder. As $1/M$ goes to 0, we
can fit, as represented on figure \ref{fig-4},
$L(M)/M$ (that also goes to 0) with a second order polynomial $a/M +b/M^2$ to
find the expected localization length $L$ equal to the parameter $a$.

A complete study of such a 2D system was recently carried out by Unge et
Stafstr\"om. The results of Ref. \onlinecite{unge} were partly used to
validate our procedure of calculation of the Lyapunov exponents.

\begin{figure}[!h]
\begin{center}
\psfrag{w}[][]{Localization length, $L/a$} 
 \includegraphics{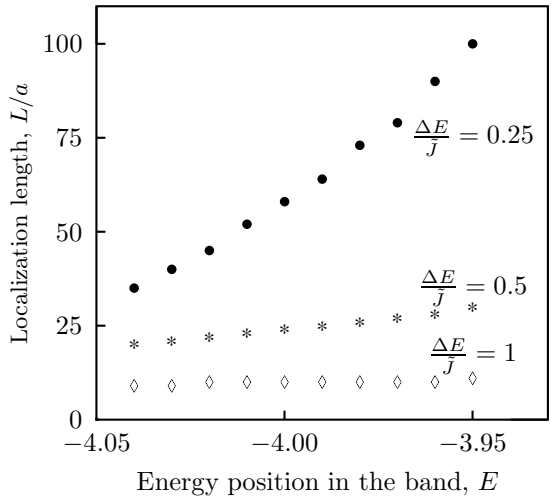}

%
%
%
%
%
%
%
%
%
%
\end{center}

\caption{Localization length close to the hole band-edge for different values of
the
diagonal disorder. When the disorder increases, the localization length
decreases.}
\label{fig-1}
\end{figure}

Since we are interested in charge transport, we explore the behavior of the
localization length near the band edge. Figure \ref{fig-1} represents the
results for an energy range between $-3.95J$ and $-4.05J$  with the band-edge of
the perfect 2D lattice being equal to $-4J$.

\section{The evaluation of the mobility}

At time scales of a fraction of picosecond or less, intermolecular phonons can be viewed as static disorder; then, at this time scale, quantum coherence induces carrier localization.
 A carrier placed in such a system extends coherently over a characteristic length
$L$. The question is  how can this electron that is localized by thermal
disorder move to a new position? If all types of intramolecular phonons
wether acoustic, optical and librational, are able to localize the charge
efficiently, they cannot drive the motion of the charge with the same
efficiency.

Dispersionless optical phonons are not efficient in moving charges. Emin was
the first to observe this fact \cite{emin} which was established more
rigorously in Ref. \onlinecite{pastawski2}. Furthermore we consider the
charge
motion to be adiabatic, because transfer integrals are at least one order of
magnitude larger than intermolecular phonon energies. The most efficient
process for moving the extra-charge is thus to use an acoustic phonon moving in
the conduction plane. Then, during the lifetime of the localized excitation,
the charge follows the phonon packet adiabatically. It moves along with the
acoustic phonon with a speed that is basically equal to the sound velocity
$v_{\text{s}}$. Values of the sound velocity in naphtalene are given in
Ref. \onlinecite{gosar} and can also be deduced from the phonon dispersion
curves of Ref. \onlinecite{natkaniec}. We have taken a value of
$v_{\text{s}}=3.3\text{km.s}^{-1}$ , which we consider a good order of magnitude for acenes.

Thus the decoherence from the Anderson localized state to the classical
diffusion in the plane, leads to a mobility

\begin{equation}
\mu=\f{|e|D}{k_{\text{B}}T}=\f{|e|v_{\text{s}}L}{4k_{\text{B}}T}
\label{eq-2}
\end{equation}

\begin{figure}[!h]

\begin{center}
\psfrag{w}[][]{Localization length, $\log(L/a)$} 
 \includegraphics{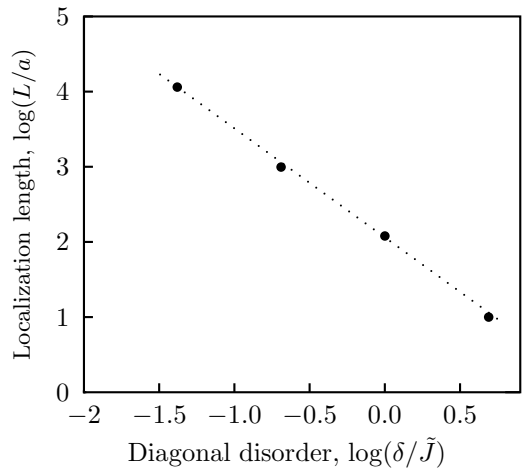}
%
%
%
%
%
%
%
%
\end{center}
\caption{Neper logarithm of the localization length calculated at
$E=-4J$ versus the Neper logarithm of the diagonal disorder. The linear
relation leads to a power-law behaviour as in experiments. The slope is about
$-1.4$.}
\label{fig-2}
\end{figure}

The Einstein's relation used here is valid for a non-degenerated hole gaz. This
condition is achieved in time of flight measurements and in
field effect
transistors. From the transfer matrix calculation applied near the hole band
edge, we find a localization length $L/a$ varying as a power law of the
polarization energy fluctuation (cf Fig. \ref{fig-2})

\begin{equation}
\f{L}{a}=8\left(\f{\delta}{\tilde{J}}\right)^{-1.4}
\end{equation}

\noindent where $a=5\text{ nm} $ is the lattice spacing parameter and $\tilde{J}=58 \text{
meV}$ the renormalized transfer integral.

If we combine the effect of a Gaussian translational disorder with a root mean
square amplitude of $0.1\text{\AA}$ and a gaussian librational disorder of 3
degrees around each axis, we get from table \ref{tab-1} a resulting energetic
disorder of $45\text{meV}$. From Figure \ref{fig-2}, we deduce a localization
length of $11a$. Using Eq. \ref{eq-2}, we get a mobility of
$1.8\text{ cm}^2.\text{V}^{-1}.\text{s}^{-1}$


Furthermore, because the polarization coupling is linear in the different
molecular degrees
of freedom (Eq. \ref{eq-3}), the fluctuation $\delta$ is proportional to
$(k_{\text{B}}T)^{1/2}$ at high temperatures. Then, we find that the mobility
varies as a power law of the temperature with exponent $\alpha\simeq-1.7$. 

It is important to have an idea of the temperature range of validity of the power laws that results from our theory. They are based on a diffusion-like view of the transport. 
The disorder characterized by $\delta$ is a site disorder, whereas the localized wavefunction spans multiple sites and is therefore characterized by a disorder that is lower by a factor of $(a/L)^2$.
%
The condition $\delta(a/L)^2<k_BT$ guarantees the diffusive aspect of the motion; otherwise, for cases in which the effective disorder exceeds $k_BT$, a hopping process appears. With the values of Table \ref{tab-1}, concerning thermal disorder and polarization fluctuations this condition yields $T>50$ K.

\section{Comparison with experiment}

The present theoretical work is applicable to the evaluation of the intrinsic mobility of a carrier, which propagates through a single crystalline molecular semiconductor such as pentacene or rubrene. In particular, it is not directly applicable to the channel of a molecular transistor with an oxide gate. In these transistors, surface Fr\"ohlich polarons have been shown to play a major role. \cite{hocine,hulea} Moreover, the field-effect mobility is affected by the presence of traps, particularly at the gate interface. Charge carrier traps dominate transport in thin film transistors and often in single crystalline transistors. \cite{podzorov2} These traps have not been considered here.

The temperature range of interest is between 50 and 400 K where the mobility decreases with increasing temperature with a power law, $T^{-\alpha}$.
The relevant experiments were performed on large ultrapurified molecular crystals by using the time-of-flight technique. 
The exponent $\alpha$ was found to be 2.9 for holes and 1.4 for electrons in naphtalene,\cite{karl2,warta} 1.5 for holes and 1.26 for electrons in anthracene,\cite{karl3} and 1.87 for electrons in perylene.\cite{karl}
%
%
%
More recently the measured values of $\alpha$ in sublimation grown perylene single crystals were found to be $2.8$ for electrons and close to zero for holes. \cite{kotani} In biphenyl single crystals, the electron mobility exponent is 1.18 for both electrons and holes. \cite{karl} In phenenthrene, the hole mobility varies with the exponent 1.8 and the electron mobility with the exponent 0.95. \cite{karl} In tetracene single crystals, an exponent $\alpha$ of 2 was reported for holes \cite{boer}, while for air-gap transistor in rubrene a value of 1.45 can be deduced above 240 K from the curve of Fig. 3 in Ref. \onlinecite{podzorov2}.

All the experiments cited above demonstrate that in ultrapure single crystals of molecular semiconductors, the mobility decreases as a power law with increasing temperature. The precise value of the exponent depends on the details of the crystal structure, on the transfer integrals, on the polarizabilities of the molecules, and on the interaction of the extra charge with the lattice. The exponent $\alpha$ varies essentially between 1 and 2 depending on these factors.

In this sense our theoretical model, which contains no adjustable parameter and describes a prototypical crystalline molecular semiconductor is in excellent agreement with experiment.
The model can be easily adapted to calculate $\alpha$ in many systems by entering the relevant material-specific parameters. The geometry of the lattice can also be chosen according to more precise crystallographic data (in the present analysis, the conducting plane of acenes is assumed to be a square molecular lattice). The values of the relevant transfer integrals and their number can also be varied (here we have considered a single transfer integral). In general, these values are different for electrons and holes. The calculation of thermal disorder at a given temperature can also be varied according to the polarizabilities of the individual molecules and the characteristics of the coupling to the lattice. Moreover, the presence of defects and impurities acting as traps in the actual crystals can always change the temperature dependence of the mobility, even for temperatures very outside of the hopping regime. Only intrinsic effects have been included in the present calculation.

It is worth summarizing here the different parameters that we have used in the model and recalling their origin. 
%
Most of these values concern pentacene, but in the cases where the values does not exist for this compound, we can also infer the value from experiments on other acenes.


 -- bare transfer integral : $\sim100$ meV from Refs. \onlinecite{cheng} and \onlinecite{troisi2}

-- effective transfer integral renormalized twice according to Ref. \onlinecite{hocine},  $\tilde{J}= 58$ meV

-- relevant intramolecular phonon frequency:\cite{coropceanu} 1340 cm$^{-1}$

-- typical intermolecular phonon frequencies:\cite{hannewald2, natkaniec} 50 cm$^{-1}$

-- sound velocity in the plane: \cite{gosar,natkaniec} 3,3 km/s

-- amplitude of the translational vibrations at room temperature:\cite{cruickshank}  
0,17\AA

-- amplitude of the librations at room temperature:\cite{cruickshank} $\sim3$ degrees

-- polarization energy of a carrier $Ep = - 1.5$ eV, calculated in Ref. \onlinecite{hocine}.


Special cases have also been observed where the mobility is relatively constant over a large temperature range. In field effect transistors this is now attributed to the existence of Fr\"ohlich polarons. \cite{hulea} In the present form our model cannot account for such a low value of $\alpha$ (close to zero).

\section{Comparison with recent theories}

The present theory depicts the behavior of a carrier in a random medium resulting from lattice fluctuations. It is important to compare it to other existing models that pursue the same purpose by using other scenarios: polarons, reorganization energies, etc.

In our model, lattice fluctuations are coupled to the charge carrier through fluctuations of the Coulomb polarization energy. \cite{gosar,mnb} The idea is to consider possible dynamic localization processes induced by these Coulomb fluctuations acting on the carrier as a random field. This means that at short time scales, the motion of the carrier is coherent and can be treated quantum-mechanically using the transfer matrix formalism. At this stage we avoid any averaging of the coupling energies. We keep quantum interference and Anderson localization, which are relevant to this problem. A study of Anderson localization in molecular semiconductors using the same transfer matrix formalism was recently published by Unge and Stafstr\"om. \cite{unge2} The hamiltonian  they use is the same as Eq. \ref{eq-5} and comes from Ref. \onlinecite{mnb}. Our results are consistent with theirs, but the authors do not follow the same analysis or present a transport model.

We consider the localization process to be dynamical. At time scales longer than 1 ps (see Appendix \ref{time}) decoherence occurs. Then the localized carrier "surfs" adiabatically on the acoustic phonon waves: in fact, the carrier localizes in a certain landscape but the landscape moves slowly and the carrier is forced to follow this motion.


In some respect, this picture resembles the scenario proposed by Troisi and Orlandi \cite{troisi} to achieve the same purpose. Our model is 2D instead of theirs which is 1D and we use quantum transport instead of a semiclassical simulation. Both models insist on the fact that lattice fluctuations can both localize the carrier and drive its diffusion in the lattice.

It is also important at this stage to compare our theory with well established recent models based on reorganization energies and polarons.

In these last years, Marcus theory has been extensively used to deduce charge transport parameters in molecular organic semiconductors. The model has been established half a century ago to predict charge transfer rates between a reactant and a product in a donor-acceptor reaction. More recently it has been widely used to determine small polaron hopping rates in oxides such as chromia, iron oxide, anatase and rutile, \cite{rosso} and in molecular organic semiconductors. \cite{gruhn,coropceanu,silvafilho,mastorrent} The advantages of this model are that the transport parameters are considered to depend only on a pair of adjacent sites  and that the carrier is coupled to the lattice through optical or intramolecular vibration modes only. The result is simple and the charge transfer rate $k_{ij}$ can be written in the semiclassical and nonadiabatic approximation :

\begin{equation}
k_{ij}={t_{ij}}^2\sqrt{\dfrac{\pi}{\hbar^2k_BT\lambda_{ij}}}\exp\left( -\dfrac{(\Delta E_{ij}-\lambda_{ij})^2}{4\lambda_{ij}k_BT}\right)
\label{eq-x}
\end{equation}

\noindent where $\lambda_{ij}$ is the reorganization energy, $\Delta E_{ij} = \varepsilon_i- \varepsilon_j$, $\varepsilon_i$ and $\varepsilon_j$ are the energies of the initial and final states, and $t_{ij}$ is the transfer integral between the two sites.

The value of $\lambda_{ij}$ corresponds to the dimer energy difference between the situation where the pair is charged but the geometric configuration corresponds to a neutral pair, and the situation where the pair is charged and is in the true geometry.

It is worth noting that, in this form, Marcus theory is strictly equivalent to the Emin-Holstein's earlier model of small polaron hopping.\cite{emin} Emin and Holstein introduced the concept of polaron binding energy $E_B$ (equal to one quarter of the reorganization energy) and the idea of coincidence, which is equivalent to the idea of a transition state in Marcus theory.

Due to the fact that Marcus theory  always yields a transfer rate that increases with temperature, it cannot adequately describe the charge mobility in the bulk of single crystals where the temperature behavior is just the opposite.
%
Even in the hopping regime, relation \ref{eq-x} has many restrictions. For example, this equation is only applicable when $\lambda/4$ is much larger than $t_{ij}$, a condition that is not satisfied for acenes and related compounds. \cite{coropceanu} It is also only valid when the relevant intramolecular phonon frequency responsible for charge transfer (1340 cm$^{-1}$) is much lower than the transfer integral $t_{ij}$. This is also not the case in acenes and related compounds. \cite{coropceanu} A third restriction is that small polaron hopping in a crystal must obey selection rules. \cite{emin, pastawski2} The optical modes involved in the small polaron formation cannot lead to carrier motion in the lattice. Acoustic phonons are mandatory for this hopping process to occur. This solid state effect is not accounted for in Marcus theory.

Finally relation \ref{eq-x} includes only short range interactions which are usually deduced from ab-initio software packages. As emphasized in reference \onlinecite{valeev}, "the polarization effect in these systems is largely electrostatic in nature and can change dramatically upon transition from a dimer to an extended system."

In order to overcome the problem of nonlocality and to introduce the low energy acoustic phonons that are of paramount importance in charge transfer processes in molecular semiconductors, Hannewald and Bobbert have worked on the basis of a Peierls-Holstein hamiltonian. Both the carrier and the phonons are treated quantum mechanically. The role of low energy phonons is to modulate the carrier energy on each site and the transfer integrals between adjacent sites. At finite temperature this modulation due to a large number of phonons modes could induce Anderson-like localization on the carrier, especially in anisotopical electron systems like acenes. However, as already observed by Troisi and Orlandi, the treatment of Hannewald and Bobbert excludes this possibility by averaging the electronic energies at all time scales. \cite{troisi}

\section{Polarization energy fluctuations or transfer integral fluctuations ?}

In general, thermal energetic disorder in organic molecular semiconductors comes either from
a distribution of polarization energies or of transfer integrals or a
combination of both of these fluctuations. Troisi and Orlandi
\cite{troisi,troisi2} have shown that by an appropriate choice for the values of
the parameters in their model, transfer integral fluctuations alone are able to
account for the order of magnitude of the mobility and its observed temperature
dependence. Here, we have also shown that polarization fluctuations alone can
achieve the same result.

In fact, the intermolecular potentials contain both short range and long range
contributions due to the presence of an extra charge. It is typical in quantum
chemistry and molecular dynamics calculations \cite{troisi,troisi2} that
only the short range part of these potentials are considered. However, long
range polarization effects are not negligible and must be included in any
realistic calculation.


It is important to emphasize the result of two recent experiments which show
that polarization contributions cannot be avoided and do not represent special
cases. The measurements of Morpurgo et al.
\cite{stassen} have shown unambiguously that the hole field effect mobilities in
single crystals of rubrene or tetracene depend strongly on the dielectric
permittivity of the gate. We have recently shown \cite{hocine} that such
dependence can be understood only if polarization effects are taken into
account.
Our recent thermopower measurements \cite{adrian} on high quality
pentacene films evaporated on different substrates revealed a large intrinsic
temperature independent,
contribution to the Seebeck coefficient of 265$\mu$V/K. This implies that each
carrier
transports an intrinsic vibrational entropy of 3 Boltzmann constants
($3k_{\text{B}}$). We have shown quantitatively that this large entropy is
associated with local variations of the low energy vibration frequencies around
a carrier by $30\%$ with respect to the bulk. This fact, which is not considered
in Ref. \onlinecite{troisi}, is easily understood when one considers that
polarization effects induced by the carrier locally change the character of the
interaction between molecules from Van de Waals bonds to point-dipole or
dipole-dipole interactions. They validate the concept of polarization
fluctuations which was introduced initially by Gosar and Choi.\cite{gosar}

In fact, the role of long range polarization in the motion of carriers in acenes and related compounds was adressed for the first time in 1954 \cite{toyozawa,gosar} and considered seriously in the seventies. \cite{silinsh} Nowadays it tends once more to be widely recognized in the scientific community. \cite{valeev,unge2,coropceanu2}

\section{Conclusion}

Following the work of Gosar and Choi, \cite{gosar} we have proposed a model that accounts for the main features of transport in semiconducting acenes and related compounds. Like Troisi and Orlandi, \cite{troisi} we think that charge transport is due to lattice fluctuations. An important contribution of the present model is the inclusion of quantum interference effects, which are generally ignored in other transport models.

The present theory exclusively concerns conduction in bulk ultra pure crystals. It is also likely to apply to conduction in ultrapure crystal transistors made with  polymer or air-gap gates. Although Fr\"ohlich polarons were not considered in the present work, we have proposed that they dominate the transport \cite{hocine} in cases where a molecular single crystal is interfaced with an oxide gate.  This conclusion is consistent with recent experiments. \cite{hulea}

Even in single crystals, the presence of electroactive defects acting as traps has been demonstrated. \cite{podzorov2} They change the temperature dependence of the mobility. In the case where these traps dominate the transport, typically in thin-film acene transistors, a hopping regime sets in and the mobility becomes essentially activated with temperature. Even in this regime, the electronic polarization induced around trapped and free charges remains of paramount importance. A work is in progress to demonstrate this point theoretically.

\appendix

\section{Time scales}
\label{time}

The non-interacting band properties of a perfect pentacene crystal
along all crystallographic directions have been calculated by
Troisi et Orlandi\cite{troisi2} and Cheng et al.\cite{cheng} For the bare transfer integral, $J^0$,
between molecules along the direction of easy propagation in the
$\left( a,b\right) $-plane, they obtain $J^0=100\text{ meV from
which one gets }h/J^0\simeq 4\times 10^{-14}$ s as the
characteristic time for in-plane Bloch-wave formation. The
corresponding transfer time in the perpendicular $c$-axis
direction, $h/J^0_{\bot }$, is thirty times longer than in the
plane. Thus, in the presence of scattering, which substantially
reduces the Bloch-wave lifetime, the carrier motion is essentially
two-dimensional.

The above considerations allow for the classification of the
various charge carriers interactions in organic
semiconductors. For fast interactions with characteristic times
shorter than $h/J^0$, the charge can be assumed to be located on a
single molecular site. In pentacene, this is the situation
encountered during the interaction of the carrier with the
electronic polarizability of the medium or in intramolecular
charge-transfer as well as the coupling with intramolecular carbon
stretching vibrations with frequencies around $1340\text{cm}^{-1}$. Since fast interactions arise prior to the formation of
the Bloch-wave, they have the effect of dressing the charge with a
polarization cloud or a lattice deformation cloud. Slow
interactions, on the other hand, have characteristic times much
longer than $h/J^0$. They act directly on the Bloch-wave or the
localized state. Such is the case for interactions of the charge
carrier with low-energy intermolecular thermal phonons and
librations, which in many cases can be considered as static with
respect to the two-dimensional band motion $\sim 50 \text{cm}^{-1}$. These interactions
scatter the Bloch-wave or localize the electronic states when the
disorder they introduce is large enough.  An interesting discussion
of time scales can also be found in the first chapter of the book
by Silinsh and \v{C}\'{a}pek.\cite{silinsh}

Because they dress the charge with a polarization cloud or lattice
deformation, fast processes lead to a renormalization of the bare
transfer integrals $J^0$ and $J^0_{\bot }$ and consequently increase
the effective mass along all crystal directions. The case
involving electron-phonon interactions has been discussed by
several authors including Appel\cite{appel} and
Davydov.\cite{davydov} The purely electronic effects were treated
in  earlier works \cite{mnb,hocine} in which we
calculated the renormalization effect due to the electronic
polarizability in the bulk of the organic semiconductor.

\section{Spatial correlations}

\label{correlations}

Consider 2 sites $n$ and $m$. The polarization energy $E_\text{p}(n)$ depends on
the spatial configuration of the molecules that corresponds to the polarization
cloud around the site $n$. Thus, if the polarization clouds around the
given sites $n$ and
$m$ overlap, $E_\text{p}(m)$ and $E_\text{p}(n)$ are correlated.

\begin{figure}[!h]
\begin{center}
\psfrag{w}[][]{Correlation, $\sigma_{0,m}$ $[ 10^{-3} \U{eV^2}]$} 
 \includegraphics{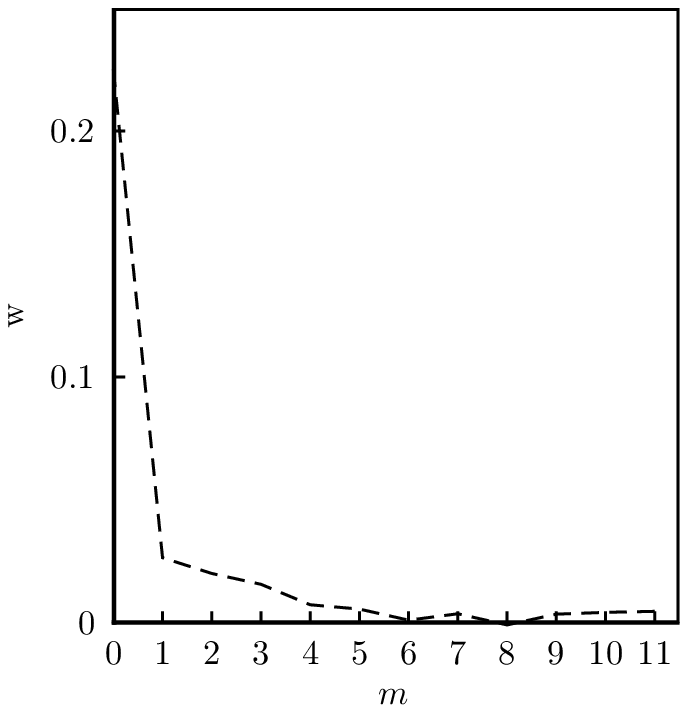}

%
%
%
\end{center}
\caption{Correlation between $E_\text{p}(0)$ and  $E_\text{p}(m)$ versus the
number of sites between site $0$ and site $m$. 
The correlation decreases very fast so
that we can neglect spatial correlations.
}
\label{fig-3}

\end{figure}

In order to evaluate the magnitude of these correlations, we calculated

\begin{equation}
\sigma_{n,m}=\lr{E_\text{p}(m)E_\text{p}(n)}-\lr{E_\text{p}(m)}\lr{E_\text{p}(n)
}
\end{equation}

\noindent for $m=n+1$ (next neighbour of $n$) to $m=n+11$ in the direction of
highest transfer integral. The discrete cluster radius was set to $18\text{
\AA}$
and the angular disorder to $5$ degrees. The results of these discrete
calculations are depicted in  Fig. \ref{fig-3}. The value for $m=0$ corresponds 
to the variance of the distribution of the polarization energy.

%
 One can readily see from these results that the correlations decrease 
 very fast with  distance to a value smaller than $10\%$ of the variance. 
This decrease is to be related to the induced character of the dipoles : permanent dipoles  would lead to long-range correlations characteristic of Coulomb interactions.
Therefore, as mentioned previously, we can reasonably neglect the spatial correlations of the polarization energy distribution and only take into account energetic correlations in the study of
the Anderson hamiltonian of Eq. \ref{eq-5}.

\section{Matrices $M_n$ and $\tilde{M}_n$ }
\label{matrices}

The transfer matrix equation (Eq. \ref{eq-9}) depends on the $2M\times2M$
matrices $M_n$
and $\tilde{M}_n$ defined as,


\begin{center}
\epsfig{file=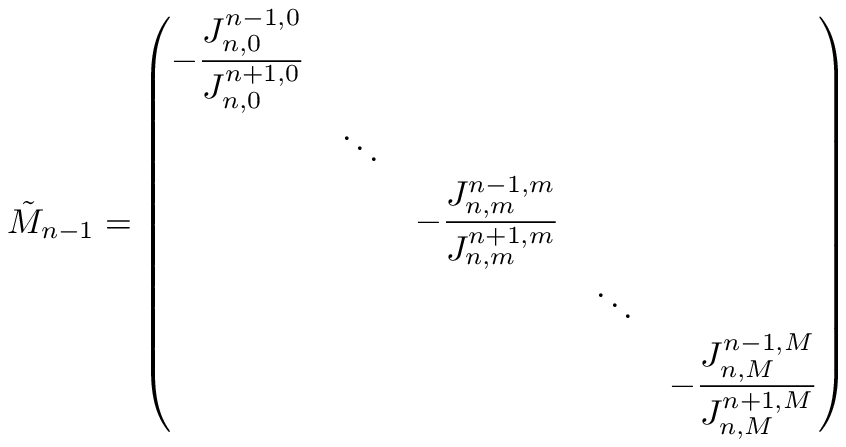}
\end{center}

\begin{widetext}

\begin{center}
\epsfig{file=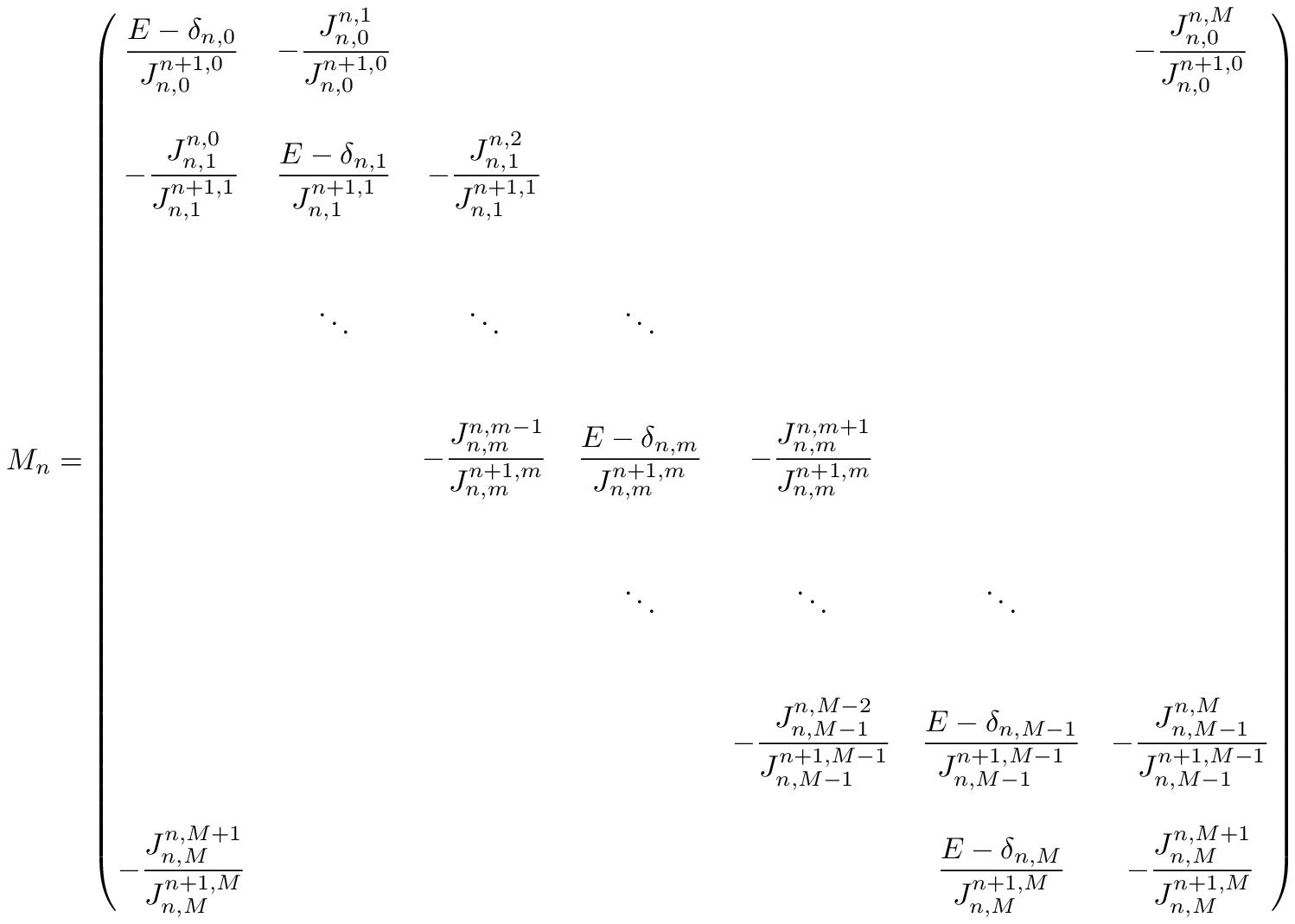, width=14.3cm}
\end{center}
\end{widetext}


\begin{acknowledgments}
The authors want to thank the maintainers of the cluster Pleiades
(\verb?pleiades.epfl.ch?) that was used heavily for our calculations. They also
acknowledge S.~J.~Konezny and D.~B.~Romero for scientific discussions and the Swiss National
Science Fundation for the financial support (project number 200020-113254).
\end{acknowledgments}

\newpage

\bibliography{biblio}

\begin{thebibliography}{46}
\expandafter\ifx\csname natexlab\endcsname\relax\def\natexlab#1{#1}\fi
\expandafter\ifx\csname bibnamefont\endcsname\relax
  \def\bibnamefont#1{#1}\fi
\expandafter\ifx\csname bibfnamefont\endcsname\relax
  \def\bibfnamefont#1{#1}\fi
\expandafter\ifx\csname citenamefont\endcsname\relax
  \def\citenamefont#1{#1}\fi
\expandafter\ifx\csname url\endcsname\relax
  \def\url#1{\texttt{#1}}\fi
\expandafter\ifx\csname urlprefix\endcsname\relax\def\urlprefix{URL }\fi
\providecommand{\bibinfo}[2]{#2}
\providecommand{\eprint}[2][]{\url{#2}}

\bibitem[{\citenamefont{Gosar and Choi}(1966)}]{gosar}
\bibinfo{author}{\bibfnamefont{P.}~\bibnamefont{Gosar}} \bibnamefont{and}
  \bibinfo{author}{\bibfnamefont{S.-I.} \bibnamefont{Choi}},
  \bibinfo{journal}{Physical Review} \textbf{\bibinfo{volume}{150}},
  \bibinfo{pages}{529} (\bibinfo{year}{1966}).

\bibitem[{\citenamefont{Bussac et~al.}(2004)\citenamefont{Bussac, Picon, and
  Zuppiroli}}]{mnb}
\bibinfo{author}{\bibfnamefont{M.-N.} \bibnamefont{Bussac}},
  \bibinfo{author}{\bibfnamefont{J.-D.} \bibnamefont{Picon}}, \bibnamefont{and}
  \bibinfo{author}{\bibfnamefont{L.}~\bibnamefont{Zuppiroli}},
  \bibinfo{journal}{Europhysics Letters} \textbf{\bibinfo{volume}{66}},
  \bibinfo{pages}{3} (\bibinfo{year}{2004}).

\bibitem[{\citenamefont{Troisi and Orlandi}(2006{\natexlab{a}})}]{troisi}
\bibinfo{author}{\bibfnamefont{A.}~\bibnamefont{Troisi}} \bibnamefont{and}
  \bibinfo{author}{\bibfnamefont{G.}~\bibnamefont{Orlandi}},
  \bibinfo{journal}{J. Phys. Chem. A} \textbf{\bibinfo{volume}{110}},
  \bibinfo{pages}{4065} (\bibinfo{year}{2006}{\natexlab{a}}).

\bibitem[{\citenamefont{Kostin et~al.}(1978)\citenamefont{Kostin, Savelev, and
  Vannikov}}]{kostin}
\bibinfo{author}{\bibfnamefont{A.~A.} \bibnamefont{Kostin}},
  \bibinfo{author}{\bibfnamefont{V.~V.} \bibnamefont{Savelev}},
  \bibnamefont{and} \bibinfo{author}{\bibfnamefont{A.~V.}
  \bibnamefont{Vannikov}}, \bibinfo{journal}{Phys. Stat. Sol. B}
  \textbf{\bibinfo{volume}{87}}, \bibinfo{pages}{255} (\bibinfo{year}{1978}).

\bibitem[{\citenamefont{Karl et~al.}(1999)\citenamefont{Karl, Kraft,
  Marktanner, Munch, Schatz, Stehle, and Unde}}]{karl}
\bibinfo{author}{\bibfnamefont{N.}~\bibnamefont{Karl}},
  \bibinfo{author}{\bibfnamefont{K.~H.} \bibnamefont{Kraft}},
  \bibinfo{author}{\bibfnamefont{J.}~\bibnamefont{Marktanner}},
  \bibinfo{author}{\bibfnamefont{M.}~\bibnamefont{Munch}},
  \bibinfo{author}{\bibfnamefont{F.}~\bibnamefont{Schatz}},
  \bibinfo{author}{\bibfnamefont{R.}~\bibnamefont{Stehle}}, \bibnamefont{and}
  \bibinfo{author}{\bibfnamefont{H.-M.} \bibnamefont{Unde}},
  \bibinfo{journal}{J. Vac. Si. Technol. A} \textbf{\bibinfo{volume}{17}},
  \bibinfo{pages}{2318} (\bibinfo{year}{1999}).

\bibitem[{\citenamefont{Warta and Karl}(1985)}]{warta}
\bibinfo{author}{\bibfnamefont{W.}~\bibnamefont{Warta}} \bibnamefont{and}
  \bibinfo{author}{\bibfnamefont{N.}~\bibnamefont{Karl}},
  \bibinfo{journal}{Phys. Rev. B} \textbf{\bibinfo{volume}{32}},
  \bibinfo{pages}{1172} (\bibinfo{year}{1985}).

\bibitem[{\citenamefont{Karl}(2003)}]{karl2}
\bibinfo{author}{\bibfnamefont{N.}~\bibnamefont{Karl}},
  \bibinfo{journal}{Synthetic Metals} \textbf{\bibinfo{volume}{133-134}},
  \bibinfo{pages}{649} (\bibinfo{year}{2003}).

\bibitem[{\citenamefont{Podzorov et~al.}(2003)\citenamefont{Podzorov, Sysoev,
  Loginova, Pudalov, and Gershenson}}]{podzorov}
\bibinfo{author}{\bibfnamefont{V.}~\bibnamefont{Podzorov}},
  \bibinfo{author}{\bibfnamefont{S.~E.} \bibnamefont{Sysoev}},
  \bibinfo{author}{\bibfnamefont{E.}~\bibnamefont{Loginova}},
  \bibinfo{author}{\bibfnamefont{V.~M.} \bibnamefont{Pudalov}},
  \bibnamefont{and} \bibinfo{author}{\bibfnamefont{M.~E.}
  \bibnamefont{Gershenson}}, \bibinfo{journal}{Appl. Phys. Lett.}
  \textbf{\bibinfo{volume}{83}}, \bibinfo{pages}{3504} (\bibinfo{year}{2003}).

\bibitem[{\citenamefont{Stassen et~al.}(2004)\citenamefont{Stassen, de~Boer,
  Iosad, and Morpurgo}}]{stassen}
\bibinfo{author}{\bibfnamefont{A.}~\bibnamefont{Stassen}},
  \bibinfo{author}{\bibfnamefont{R.~W.~I.} \bibnamefont{de~Boer}},
  \bibinfo{author}{\bibfnamefont{N.~N.} \bibnamefont{Iosad}}, \bibnamefont{and}
  \bibinfo{author}{\bibfnamefont{A.~F.} \bibnamefont{Morpurgo}},
  \bibinfo{journal}{Appl. Phys. Lett} \textbf{\bibinfo{volume}{85}},
  \bibinfo{pages}{3899} (\bibinfo{year}{2004}).

\bibitem[{\citenamefont{Gruhn et~al.}(2002)\citenamefont{Gruhn, da~Silva~Filho,
  Bill, Malagoli, Coropceanu, Kahn, and Br\'edas}}]{gruhn}
\bibinfo{author}{\bibfnamefont{N.~E.} \bibnamefont{Gruhn}},
  \bibinfo{author}{\bibfnamefont{D.~A.} \bibnamefont{da~Silva~Filho}},
  \bibinfo{author}{\bibfnamefont{T.~G.} \bibnamefont{Bill}},
  \bibinfo{author}{\bibfnamefont{M.}~\bibnamefont{Malagoli}},
  \bibinfo{author}{\bibfnamefont{V.}~\bibnamefont{Coropceanu}},
  \bibinfo{author}{\bibfnamefont{A.}~\bibnamefont{Kahn}}, \bibnamefont{and}
  \bibinfo{author}{\bibfnamefont{J.-L.} \bibnamefont{Br\'edas}},
  \bibinfo{journal}{J. Am. Chem. Soc.} \textbf{\bibinfo{volume}{124}},
  \bibinfo{pages}{7918} (\bibinfo{year}{2002}).

\bibitem[{\citenamefont{Coropceanu et~al.}(2002)\citenamefont{Coropceanu,
  Malagoli, da~Silva~Filho, Gruhn, Bill, and Br\'edas}}]{coropceanu}
\bibinfo{author}{\bibfnamefont{V.}~\bibnamefont{Coropceanu}},
  \bibinfo{author}{\bibfnamefont{M.}~\bibnamefont{Malagoli}},
  \bibinfo{author}{\bibfnamefont{D.~A.} \bibnamefont{da~Silva~Filho}},
  \bibinfo{author}{\bibfnamefont{N.~E.} \bibnamefont{Gruhn}},
  \bibinfo{author}{\bibfnamefont{T.~G.} \bibnamefont{Bill}}, \bibnamefont{and}
  \bibinfo{author}{\bibfnamefont{J.-L.} \bibnamefont{Br\'edas}},
  \bibinfo{journal}{Phys. Rev. Lett.} \textbf{\bibinfo{volume}{89}},
  \bibinfo{pages}{275503} (\bibinfo{year}{2002}).

\bibitem[{\citenamefont{da~Silva~Filho
  et~al.}(2005)\citenamefont{da~Silva~Filho, Kim, and Br\'edas}}]{silvafilho}
\bibinfo{author}{\bibfnamefont{D.~A.} \bibnamefont{da~Silva~Filho}},
  \bibinfo{author}{\bibfnamefont{E.-G.} \bibnamefont{Kim}}, \bibnamefont{and}
  \bibinfo{author}{\bibfnamefont{J.-L.} \bibnamefont{Br\'edas}},
  \bibinfo{journal}{Advanced Materials} \textbf{\bibinfo{volume}{17}},
  \bibinfo{pages}{1072} (\bibinfo{year}{2005}).

\bibitem[{\citenamefont{Mas-Torrent et~al.}(2004)\citenamefont{Mas-Torrent,
  Hadley, Bromley, Ribas, Torr\'es, Mas, Molins, Veciana, and
  Rovira}}]{mastorrent}
\bibinfo{author}{\bibfnamefont{M.}~\bibnamefont{Mas-Torrent}},
  \bibinfo{author}{\bibfnamefont{P.}~\bibnamefont{Hadley}},
  \bibinfo{author}{\bibfnamefont{S.~T.} \bibnamefont{Bromley}},
  \bibinfo{author}{\bibfnamefont{X.}~\bibnamefont{Ribas}},
  \bibinfo{author}{\bibfnamefont{J.}~\bibnamefont{Torr\'es}},
  \bibinfo{author}{\bibfnamefont{M.}~\bibnamefont{Mas}},
  \bibinfo{author}{\bibfnamefont{E.}~\bibnamefont{Molins}},
  \bibinfo{author}{\bibfnamefont{J.}~\bibnamefont{Veciana}}, \bibnamefont{and}
  \bibinfo{author}{\bibfnamefont{C.}~\bibnamefont{Rovira}},
  \bibinfo{journal}{J. Am. Chem. Soc.} \textbf{\bibinfo{volume}{126}},
  \bibinfo{pages}{8546} (\bibinfo{year}{2004}).

\bibitem[{\citenamefont{Kenkre et~al.}(1989)\citenamefont{Kenkre, Andersen,
  Dunlap, and Duke}}]{kenkre}
\bibinfo{author}{\bibfnamefont{V.}~\bibnamefont{Kenkre}},
  \bibinfo{author}{\bibfnamefont{J.}~\bibnamefont{Andersen}},
  \bibinfo{author}{\bibfnamefont{D.}~\bibnamefont{Dunlap}}, \bibnamefont{and}
  \bibinfo{author}{\bibfnamefont{C.}~\bibnamefont{Duke}},
  \bibinfo{journal}{Phys. Rev Lett.} \textbf{\bibinfo{volume}{62}},
  \bibinfo{pages}{1165} (\bibinfo{year}{1989}).

\bibitem[{\citenamefont{Hannewald and Bobbert}(2004)}]{hannewald}
\bibinfo{author}{\bibfnamefont{K.}~\bibnamefont{Hannewald}} \bibnamefont{and}
  \bibinfo{author}{\bibfnamefont{P.}~\bibnamefont{Bobbert}},
  \bibinfo{journal}{Phys. Rev. B} \textbf{\bibinfo{volume}{69}},
  \bibinfo{pages}{075212} (\bibinfo{year}{2004}).

\bibitem[{\citenamefont{Hannewald et~al.}(2004)\citenamefont{Hannewald,
  Stojanovi\'c, Schellekens, and Bobbert}}]{hannewald2}
\bibinfo{author}{\bibfnamefont{K.}~\bibnamefont{Hannewald}},
  \bibinfo{author}{\bibfnamefont{V.}~\bibnamefont{Stojanovi\'c}},
  \bibinfo{author}{\bibfnamefont{J.}~\bibnamefont{Schellekens}},
  \bibnamefont{and} \bibinfo{author}{\bibfnamefont{P.}~\bibnamefont{Bobbert}},
  \bibinfo{journal}{Phys. Rev. B} \textbf{\bibinfo{volume}{69}},
  \bibinfo{pages}{075211} (\bibinfo{year}{2004}).

\bibitem[{\citenamefont{Troisi and Orlandi}(2006{\natexlab{b}})}]{troisi2}
\bibinfo{author}{\bibfnamefont{A.}~\bibnamefont{Troisi}} \bibnamefont{and}
  \bibinfo{author}{\bibfnamefont{G.}~\bibnamefont{Orlandi}},
  \bibinfo{journal}{Phys. Rev. Lett.} \textbf{\bibinfo{volume}{96}},
  \bibinfo{pages}{086601} (\bibinfo{year}{2006}{\natexlab{b}}).

\bibitem[{\citenamefont{Johansson and
  Stafstr$\Ddot{\text{o}}$m}(2004)}]{johansson}
\bibinfo{author}{\bibfnamefont{A.~A.} \bibnamefont{Johansson}}
  \bibnamefont{and}
  \bibinfo{author}{\bibfnamefont{S.}~\bibnamefont{Stafstr$\Ddot{\text{o}}$m}},
  \bibinfo{journal}{Phys. Rev. B} \textbf{\bibinfo{volume}{69}},
  \bibinfo{pages}{235205} (\bibinfo{year}{2004}).

\bibitem[{\citenamefont{Natkaniec et~al.}(1980)\citenamefont{Natkaniec,
  Bokhenkov, Dorner, Kalus, Mackenzie, Pawley, Schnelzer, and
  Sheka}}]{natkaniec}
\bibinfo{author}{\bibfnamefont{I.}~\bibnamefont{Natkaniec}},
  \bibinfo{author}{\bibfnamefont{E.~L.} \bibnamefont{Bokhenkov}},
  \bibinfo{author}{\bibfnamefont{B.}~\bibnamefont{Dorner}},
  \bibinfo{author}{\bibfnamefont{J.}~\bibnamefont{Kalus}},
  \bibinfo{author}{\bibfnamefont{G.~A.} \bibnamefont{Mackenzie}},
  \bibinfo{author}{\bibfnamefont{G.~S.} \bibnamefont{Pawley}},
  \bibinfo{author}{\bibfnamefont{V.}~\bibnamefont{Schnelzer}},
  \bibnamefont{and} \bibinfo{author}{\bibfnamefont{E.~F.} \bibnamefont{Sheka}},
  \bibinfo{journal}{J. Phys. C : Sol. St. Phys} \textbf{\bibinfo{volume}{13}},
  \bibinfo{pages}{4265} (\bibinfo{year}{1980}).

\bibitem[{\citenamefont{Cruickshank}(1958)}]{cruickshank}
\bibinfo{author}{\bibfnamefont{D.~W.~J.} \bibnamefont{Cruickshank}},
  \bibinfo{journal}{Review of Modern Physics} \textbf{\bibinfo{volume}{30}},
  \bibinfo{pages}{163} (\bibinfo{year}{1958}).

\bibitem[{\citenamefont{Cheng et~al.}(2003)\citenamefont{Cheng, Silbey,
  da~Silva~Filho, Calbert, Cornil, and Br\'edas}}]{cheng}
\bibinfo{author}{\bibfnamefont{Y.}~\bibnamefont{Cheng}},
  \bibinfo{author}{\bibfnamefont{R.}~\bibnamefont{Silbey}},
  \bibinfo{author}{\bibfnamefont{D.}~\bibnamefont{da~Silva~Filho}},
  \bibinfo{author}{\bibfnamefont{J.}~\bibnamefont{Calbert}},
  \bibinfo{author}{\bibfnamefont{J.}~\bibnamefont{Cornil}}, \bibnamefont{and}
  \bibinfo{author}{\bibfnamefont{J.}~\bibnamefont{Br\'edas}},
  \bibinfo{journal}{J. Chem. Phys.} \textbf{\bibinfo{volume}{118}},
  \bibinfo{pages}{8} (\bibinfo{year}{2003}).

\bibitem[{\citenamefont{Pastawski
  et~al.}(2002{\natexlab{a}})\citenamefont{Pastawski, Foa~Torres, and
  Medina}}]{pastawski}
\bibinfo{author}{\bibfnamefont{H.~M.} \bibnamefont{Pastawski}},
  \bibinfo{author}{\bibfnamefont{L.~E.~F.} \bibnamefont{Foa~Torres}},
  \bibnamefont{and} \bibinfo{author}{\bibfnamefont{E.}~\bibnamefont{Medina}},
  \bibinfo{journal}{Chem. Phys.} \textbf{\bibinfo{volume}{281}},
  \bibinfo{pages}{257} (\bibinfo{year}{2002}{\natexlab{a}}).

\bibitem[{\citenamefont{Houili et~al.}(2006)\citenamefont{Houili, Picon,
  Bussac, and Zuppiroli}}]{hocine}
\bibinfo{author}{\bibfnamefont{H.}~\bibnamefont{Houili}},
  \bibinfo{author}{\bibfnamefont{J.-D.} \bibnamefont{Picon}},
  \bibinfo{author}{\bibfnamefont{M.-N.} \bibnamefont{Bussac}},
  \bibnamefont{and}
  \bibinfo{author}{\bibfnamefont{L.}~\bibnamefont{Zuppiroli}},
  \bibinfo{journal}{Appl. Phys.} \textbf{\bibinfo{volume}{100}},
  \bibinfo{pages}{023702} (\bibinfo{year}{2006}).

\bibitem[{\citenamefont{Pope and Swenberg}(1999)}]{pope}
\bibinfo{author}{\bibfnamefont{M.}~\bibnamefont{Pope}} \bibnamefont{and}
  \bibinfo{author}{\bibfnamefont{C.~E.} \bibnamefont{Swenberg}},
  \emph{\bibinfo{title}{Electronic Processes in Organic Crystals and Polymers}}
  (\bibinfo{publisher}{Oxford Science Publication}, \bibinfo{year}{1999}), p.
  \bibinfo{pages}{553}.

\bibitem[{\citenamefont{Song et~al.}(2002)\citenamefont{Song, Han, Chu, Kim,
  Kim, Lyapustina, Xu, Stilles, and Bowen}}]{song}
\bibinfo{author}{\bibfnamefont{J.~K.} \bibnamefont{Song}},
  \bibinfo{author}{\bibfnamefont{S.~Y.} \bibnamefont{Han}},
  \bibinfo{author}{\bibfnamefont{I.}~\bibnamefont{Chu}},
  \bibinfo{author}{\bibfnamefont{J.}~\bibnamefont{Kim}},
  \bibinfo{author}{\bibfnamefont{S.~K.} \bibnamefont{Kim}},
  \bibinfo{author}{\bibfnamefont{S.~A.} \bibnamefont{Lyapustina}},
  \bibinfo{author}{\bibfnamefont{S.}~\bibnamefont{Xu}},
  \bibinfo{author}{\bibfnamefont{J.~M.} \bibnamefont{Stilles}},
  \bibnamefont{and} \bibinfo{author}{\bibfnamefont{K.~H.} \bibnamefont{Bowen}},
  \bibinfo{journal}{J. Chem. Phys.} \textbf{\bibinfo{volume}{116}},
  \bibinfo{pages}{4477} (\bibinfo{year}{2002}).

\bibitem[{\citenamefont{Unge and Stafstr$\Ddot{\text{o}}$m}(2003)}]{unge}
\bibinfo{author}{\bibfnamefont{M.}~\bibnamefont{Unge}} \bibnamefont{and}
  \bibinfo{author}{\bibfnamefont{S.}~\bibnamefont{Stafstr$\Ddot{\text{o}}$m}},
  \bibinfo{journal}{Synthetic Metals} \textbf{\bibinfo{volume}{139}},
  \bibinfo{pages}{239} (\bibinfo{year}{2003}).

\bibitem[{\citenamefont{Kramer and MacKinnon}(1993)}]{kramer}
\bibinfo{author}{\bibfnamefont{B.}~\bibnamefont{Kramer}} \bibnamefont{and}
  \bibinfo{author}{\bibfnamefont{A.}~\bibnamefont{MacKinnon}},
  \bibinfo{journal}{Rep. Prog. Phys.} \textbf{\bibinfo{volume}{56}},
  \bibinfo{pages}{1469} (\bibinfo{year}{1993}).

\bibitem[{\citenamefont{Unge and Stafstr\"om}(2006)}]{unge2}
\bibinfo{author}{\bibfnamefont{M.}~\bibnamefont{Unge}} \bibnamefont{and}
  \bibinfo{author}{\bibfnamefont{S.}~\bibnamefont{Stafstr\"om}},
  \bibinfo{journal}{Phys. Rev. B} \textbf{\bibinfo{volume}{74}},
  \bibinfo{pages}{235403} (\bibinfo{year}{2006}).

\bibitem[{\citenamefont{Oseledec}(1968)}]{oseledec}
\bibinfo{author}{\bibfnamefont{V.}~\bibnamefont{Oseledec}},
  \bibinfo{journal}{Trans. Moscow Math. Soc.} \textbf{\bibinfo{volume}{19}},
  \bibinfo{pages}{197} (\bibinfo{year}{1968}).

\bibitem[{\citenamefont{Benettin and Galgani}(1979)}]{benettin}
\bibinfo{author}{\bibfnamefont{G.}~\bibnamefont{Benettin}} \bibnamefont{and}
  \bibinfo{author}{\bibfnamefont{L.}~\bibnamefont{Galgani}},
  \emph{\bibinfo{title}{Intrinsic Stochasticity in Plasmas}}
  (\bibinfo{publisher}{\'Editions de Physique, Orsay}, \bibinfo{year}{1979}),
  p.~\bibinfo{pages}{93}.

\bibitem[{\citenamefont{MacKinnon and Kramer}(1981)}]{mackinnon}
\bibinfo{author}{\bibfnamefont{A.}~\bibnamefont{MacKinnon}} \bibnamefont{and}
  \bibinfo{author}{\bibfnamefont{B.}~\bibnamefont{Kramer}},
  \bibinfo{journal}{Phys. Rev. Lett.} \textbf{\bibinfo{volume}{47}},
  \bibinfo{pages}{21} (\bibinfo{year}{1981}).

\bibitem[{\citenamefont{Emin}(1973)}]{emin}
\bibinfo{author}{\bibfnamefont{D.}~\bibnamefont{Emin}},
  \emph{\bibinfo{title}{Properties of Amorphous Semiconductors}}
  (\bibinfo{publisher}{P.G. Le Comber and J. Mort, Academic Press},
  \bibinfo{year}{1973}), pp. \bibinfo{pages}{291--293}.

\bibitem[{\citenamefont{Pastawski
  et~al.}(2002{\natexlab{b}})\citenamefont{Pastawski, Foa~Torres, and
  Medina}}]{pastawski2}
\bibinfo{author}{\bibfnamefont{H.~M.} \bibnamefont{Pastawski}},
  \bibinfo{author}{\bibfnamefont{L.~E.~F.} \bibnamefont{Foa~Torres}},
  \bibnamefont{and} \bibinfo{author}{\bibfnamefont{E.}~\bibnamefont{Medina}},
  in  \cite{pastawski}, p. \bibinfo{pages}{257}.

\bibitem[{\citenamefont{Hulea et~al.}(2006)\citenamefont{Hulea, Fratini, Xie,
  Mulder, Tossad, Rastelli, Ciuchi, and Morpurgo}}]{hulea}
\bibinfo{author}{\bibfnamefont{I.~N.} \bibnamefont{Hulea}},
  \bibinfo{author}{\bibfnamefont{S.}~\bibnamefont{Fratini}},
  \bibinfo{author}{\bibfnamefont{H.}~\bibnamefont{Xie}},
  \bibinfo{author}{\bibfnamefont{C.~F.} \bibnamefont{Mulder}},
  \bibinfo{author}{\bibfnamefont{N.~N.} \bibnamefont{Tossad}},
  \bibinfo{author}{\bibfnamefont{G.}~\bibnamefont{Rastelli}},
  \bibinfo{author}{\bibfnamefont{S.}~\bibnamefont{Ciuchi}}, \bibnamefont{and}
  \bibinfo{author}{\bibfnamefont{A.~F.} \bibnamefont{Morpurgo}},
  \bibinfo{journal}{Nature Materials} \textbf{\bibinfo{volume}{5}},
  \bibinfo{pages}{982} (\bibinfo{year}{2006}).

\bibitem[{\citenamefont{Podzorov et~al.}(2005)\citenamefont{Podzorov, Menard,
  Rogers, and Gershenson}}]{podzorov2}
\bibinfo{author}{\bibfnamefont{V.}~\bibnamefont{Podzorov}},
  \bibinfo{author}{\bibfnamefont{E.}~\bibnamefont{Menard}},
  \bibinfo{author}{\bibfnamefont{J.~A.} \bibnamefont{Rogers}},
  \bibnamefont{and} \bibinfo{author}{\bibfnamefont{M.~E.}
  \bibnamefont{Gershenson}}, \bibinfo{journal}{Phys. Rev. Lett.}
  \textbf{\bibinfo{volume}{95}}, \bibinfo{pages}{226601}
  (\bibinfo{year}{2005}).

\bibitem[{\citenamefont{Karl et~al.}(1991)\citenamefont{Karl, Marktanner,
  Stehle, and Warta}}]{karl3}
\bibinfo{author}{\bibfnamefont{N.}~\bibnamefont{Karl}},
  \bibinfo{author}{\bibfnamefont{J.}~\bibnamefont{Marktanner}},
  \bibinfo{author}{\bibfnamefont{R.}~\bibnamefont{Stehle}}, \bibnamefont{and}
  \bibinfo{author}{\bibfnamefont{W.}~\bibnamefont{Warta}},
  \bibinfo{journal}{Synthetic Metals}  (\bibinfo{year}{1991}).

\bibitem[{\citenamefont{Kotani et~al.}(2006)\citenamefont{Kotani, Kakinuma, ,
  Yoshimura, Ishii, Yamazaki, Kobori, Hokuyama, Kobayshi, and Tada}}]{kotani}
\bibinfo{author}{\bibfnamefont{M.}~\bibnamefont{Kotani}},
  \bibinfo{author}{\bibfnamefont{K.}~\bibnamefont{Kakinuma}}, ,
  \bibinfo{author}{\bibfnamefont{M.}~\bibnamefont{Yoshimura}},
  \bibinfo{author}{\bibfnamefont{K.}~\bibnamefont{Ishii}},
  \bibinfo{author}{\bibfnamefont{S.}~\bibnamefont{Yamazaki}},
  \bibinfo{author}{\bibfnamefont{T.}~\bibnamefont{Kobori}},
  \bibinfo{author}{\bibfnamefont{H.}~\bibnamefont{Hokuyama}},
  \bibinfo{author}{\bibfnamefont{H.}~\bibnamefont{Kobayshi}}, \bibnamefont{and}
  \bibinfo{author}{\bibfnamefont{H.}~\bibnamefont{Tada}},
  \bibinfo{journal}{Chemical Physics} \textbf{\bibinfo{volume}{325}},
  \bibinfo{pages}{160} (\bibinfo{year}{2006}).

\bibitem[{\citenamefont{de~Boer et~al.}(2004)\citenamefont{de~Boer, Jochemsen,
  Klapwijk, and Morpurgo}}]{boer}
\bibinfo{author}{\bibfnamefont{R.~W.~I.} \bibnamefont{de~Boer}},
  \bibinfo{author}{\bibfnamefont{M.}~\bibnamefont{Jochemsen}},
  \bibinfo{author}{\bibfnamefont{T.~M.} \bibnamefont{Klapwijk}},
  \bibnamefont{and} \bibinfo{author}{\bibfnamefont{A.~F.}
  \bibnamefont{Morpurgo}}, \bibinfo{journal}{J. Appl. Phys.}
  \textbf{\bibinfo{volume}{95}}, \bibinfo{pages}{1196} (\bibinfo{year}{2004}).

\bibitem[{\citenamefont{Rosso and Dupuis}(2004)}]{rosso}
\bibinfo{author}{\bibfnamefont{K.~M.} \bibnamefont{Rosso}} \bibnamefont{and}
  \bibinfo{author}{\bibfnamefont{M.}~\bibnamefont{Dupuis}},
  \bibinfo{journal}{Journal of Chemical Physics}
  \textbf{\bibinfo{volume}{120}}, \bibinfo{pages}{7050} (\bibinfo{year}{2004}).

\bibitem[{\citenamefont{Valeev et~al.}(2006)\citenamefont{Valeev, Coropceanu,
  da~Silva~Filho, Salman, and Br\'edas}}]{valeev}
\bibinfo{author}{\bibfnamefont{E.~F.} \bibnamefont{Valeev}},
  \bibinfo{author}{\bibfnamefont{V.}~\bibnamefont{Coropceanu}},
  \bibinfo{author}{\bibfnamefont{D.~A.} \bibnamefont{da~Silva~Filho}},
  \bibinfo{author}{\bibfnamefont{S.}~\bibnamefont{Salman}}, \bibnamefont{and}
  \bibinfo{author}{\bibfnamefont{J.~L.} \bibnamefont{Br\'edas}},
  \bibinfo{journal}{J. Am. Chem. Soc.} \textbf{\bibinfo{volume}{128}},
  \bibinfo{pages}{9882} (\bibinfo{year}{2006}).

\bibitem[{\citenamefont{von M$\Ddot{\text{u}}$hlenen
  et~al.}(2007)\citenamefont{von M$\Ddot{\text{u}}$hlenen, Errien, Schaer,
  Bussac, and Zuppiroli}}]{adrian}
\bibinfo{author}{\bibfnamefont{A.}~\bibnamefont{von M$\Ddot{\text{u}}$hlenen}},
  \bibinfo{author}{\bibfnamefont{N.}~\bibnamefont{Errien}},
  \bibinfo{author}{\bibfnamefont{M.}~\bibnamefont{Schaer}},
  \bibinfo{author}{\bibfnamefont{M.-N.} \bibnamefont{Bussac}},
  \bibnamefont{and}
  \bibinfo{author}{\bibfnamefont{L.}~\bibnamefont{Zuppiroli}},
  \bibinfo{journal}{unpublished}  (\bibinfo{year}{2007}).

\bibitem[{\citenamefont{Toyozawa}(1954)}]{toyozawa}
\bibinfo{author}{\bibfnamefont{Y.}~\bibnamefont{Toyozawa}},
  \bibinfo{journal}{Prog. Theor. Phys.} \textbf{\bibinfo{volume}{12}},
  \bibinfo{pages}{421} (\bibinfo{year}{1954}).

\bibitem[{\citenamefont{Silinsh and {\v C}\'apek}(1994)}]{silinsh}
\bibinfo{author}{\bibfnamefont{E.~A.} \bibnamefont{Silinsh}} \bibnamefont{and}
  \bibinfo{author}{\bibfnamefont{V.}~\bibnamefont{{\v C}\'apek}},
  \emph{\bibinfo{title}{Organic Molecular Crystals}} (\bibinfo{publisher}{AIP
  Press, New York}, \bibinfo{year}{1994}).

\bibitem[{\citenamefont{Coropceanu and Br\'edas}(2006)}]{coropceanu2}
\bibinfo{author}{\bibfnamefont{V.}~\bibnamefont{Coropceanu}} \bibnamefont{and}
  \bibinfo{author}{\bibfnamefont{J.~L.} \bibnamefont{Br\'edas}},
  \bibinfo{journal}{Nature Materials} \textbf{\bibinfo{volume}{5}},
  \bibinfo{pages}{929} (\bibinfo{year}{2006}).

\bibitem[{\citenamefont{Appel}(1968)}]{appel}
\bibinfo{author}{\bibfnamefont{J.}~\bibnamefont{Appel}},
  \bibinfo{journal}{Solid Stat. Phys.} \textbf{\bibinfo{volume}{25}},
  \bibinfo{pages}{193} (\bibinfo{year}{1968}).

\bibitem[{\citenamefont{Davydov}(1980)}]{davydov}
\bibinfo{author}{\bibfnamefont{A.~S.} \bibnamefont{Davydov}},
  \emph{\bibinfo{title}{Th\'eorie du solide}} (\bibinfo{publisher}{\'Editions
  Mir}, \bibinfo{year}{1980}).

\end{thebibliography}

\end{document}